\documentclass[
aps,
pra,
twocolumn,
amsmath,
amssymb,
graphicx,
bibnotes,
longbibliography
]
{revtex4-1} 

\newcommand{\bcen}{\begin{center}}
\newcommand{\ecen}{\end{center}}
\newcommand{\btab}{\begin{tabular}}
\newcommand{\etab}{\end{tabular}}
\newcommand{\bdes}{\begin{description}}
\newcommand{\edes}{\end{description}}

\newcommand{\beq}{\begin{equation}}
\newcommand{\eeq}{\end{equation}}
\newcommand{\bea}{\begin{eqnarray}}
\newcommand{\eea}{\end{eqnarray}}

\newcommand{\bary}{\begin{array}}
\newcommand{\eary}{\end{array}}
\newcommand{\benum}{\begin{enumerate}}
\newcommand{\eenum}{\end{enumerate}}
\newcommand{\bitem}{\begin{itemize}}
\newcommand{\eitem}{\end{itemize}}

\newcommand{\bk} { {\boldsymbol{k}} }
\newcommand{\br} { {\boldsymbol{r}}}

\newcommand{\bzero} { {\boldsymbol{0}}}

\newcommand{\eqn}[1] {eqn.~(\ref{#1})}

\newcommand{\Fig}[1]{Fig.~\ref{#1}}
\newcommand{\mylabel}[1]{\label{#1}}

\usepackage{bibunits}
\usepackage{hyperref}
\usepackage{amsmath}
\usepackage{epsfig}
\usepackage{wrapfig} 
\usepackage{mathrsfs}
\usepackage{amssymb} 
\usepackage{color}
\usepackage{amsbsy}
\usepackage{subfigure}
\usepackage{marginnote}
\usepackage{slashed}
\usepackage{marvosym}
\usepackage{pifont}
\usepackage{mathbbol}
\usepackage{wrapfig}
\usepackage{upgreek}

\newcommand{\usenomenclature}{}
\ifdefined\usenomenclature
\usepackage[noprefix]{nomencl}
\fi

\newcommand{\oibook}[1]{}

\newcommand{\mytitle}{
Fractalized Metals
}

\newcommand{\myaffl}{{Department of Physics, Indian Institute of Science, Bangalore 560012, India.}}
\newcommand{\myafflicts}{{International Centre for Theoretical Sciences, Tata Institute of Fundamental Research, Bangalore 560089, India.}}

\defaultbibliographystyle{unsrt}
\defaultbibliographystyle{apsrev-nourl}

\begin{document}

\title{\mytitle}
\author{Adhip Agarwala$^{1,2}$}
\author{Shriya Pai$^{1}$}
\author{Vijay B.~Shenoy$^{1}$}\email{shenoy@physics.iisc.ernet.in}
\affiliation{$^1$\myaffl \\ $^2$\myafflicts}
%\affil[2]{\myaffl}

\date{\today{}}
\begin{abstract} 
The classification of gapped phases of non-interacting fermions hinges on the tenfold symmetries and on the spatial dimension. The notion of dimension leads to a well defined demarcation between bulk and edge. Here we explore the nature of topological phases in systems where the distinction between bulk and edge is nebulous, of which fractal lattices are canonical examples. Our key finding is that in homogeneous fractal lattices (where every site is equally coordinated), there are no gapped topological phases. What appears instead is a novel metallic state -- the fractalized metal -- whose low energy states arrange hierarchically on the structure of the fractal that hosts them. We study the properties (such as chiral transport) of this metal and demonstrate its robustness to disorder. Further, by studying a variety of fractal models we establish that the  homogeneity of the fractal is a key condition for the realization of such fractalized metallic states.
\end{abstract}

%\pacs{71.10.-w, 71.27.+a, 71.10.Fd}

\maketitle 

\begin{bibunit}
%\section{Introduction}

\noindent An important contemporary development in physics is the identification and classification of gapped fermionic systems using concepts of topology and entanglement \cite{Hasan_RMP_2010, Qi_RMP_2011,Chiu_RMP_2016, Ludwig_PS_2015}.
This field has witnessed remarkable success for such phases in crystalline systems, in any spatial dimension \cite{Kitaev_AIP_2009, Schnyder_PRB_2008}. A particularly interesting, and even useful aspect, of topologically nontrivial (short range entangled) phases is the presence of robust (not affected by weak disorder) states on the boundaries of a finite system hosting a bulk topological phase. Recently it has become clearer that such topological phases needs only the notion of spatial dimension, in that they can be realized, even in amorphous lattices \cite{Adhip_PRL_2017, Mitchell_NP_2018}, which only preserves the notion of a ``bulk'' and an ``edge".

\begin{figure}
	\includegraphics[width=\columnwidth]{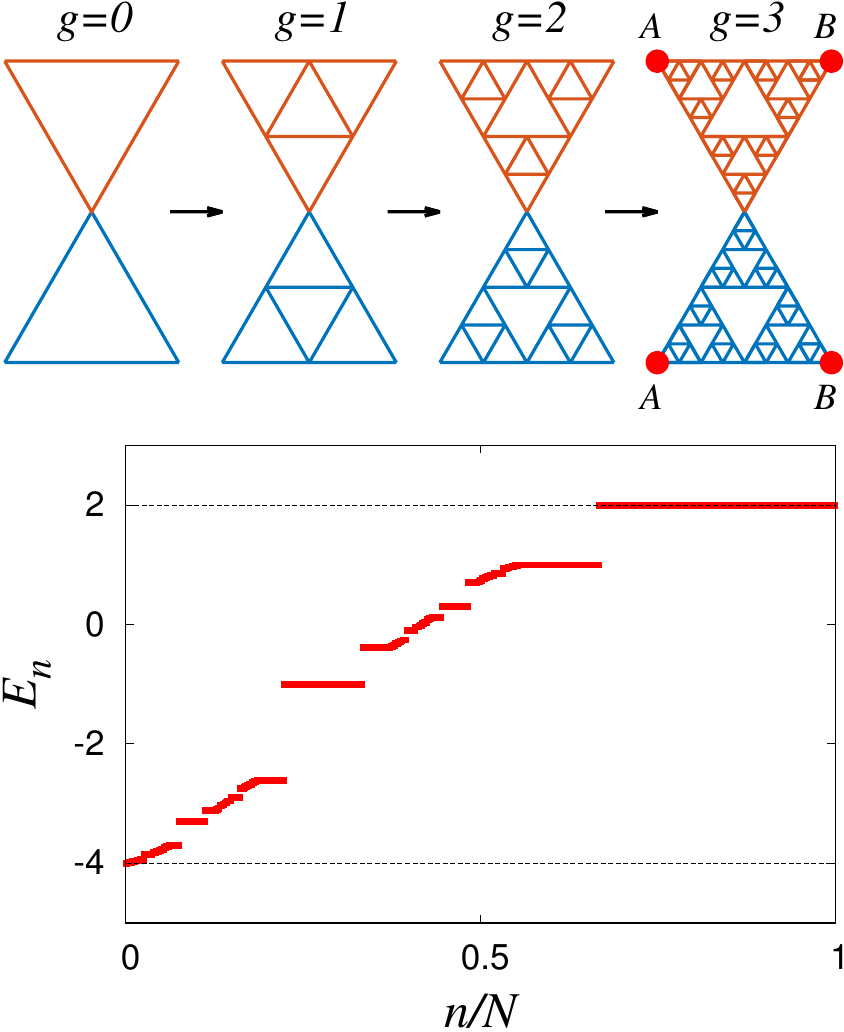}
	\caption{{\bf The Sierpinski gasket:} (top) The Sierpinski gasket on a torus is shown for different generations. $A$ sites are identified with each other, as are $B$. (Bottom) The eigenvalues as a function of normalized eigenvalue number for a simple tight binding model with unit hopping $(-t, |t|=1)$ defined on this system for $g=5$.  In infinite $g$ limit the spectrum is infinitely gapped and is self similar \cite{Domany_PRB_1983}.}
	\label{SierTB}
\end{figure}

\noindent An interesting question to explore is apropos the notion of topological phases in lattice systems which do not have a natural demarcation of a bulk and an edge. This will add to our fundamental understanding of topological phases. A fractal lattice provides a natural setting to investigate this question. A fractal lattice \cite{Mandelbrot_Book_1983} is formed by a set of sites finitely coordinated by ``neighboring" sites (``neighbor" defined on a microscopic scale) such that the system on a large scale is characterized by a fractional dimension -- called the Hausdorff dimension \cite{Mandelbrot_Book_1983}. Examples include Sierpinski gasket, Sierpinski carpet, Koch curve etc.~\cite{Mandelbrot_Book_1983}. Physical phenomena  realized on fractal lattices have intrigued physicists from various areas including material science \cite{Alexander_JPL_1982, Domany_PRB_1983}, statistical physics of phase transitions\cite{Gefen_PRL_1980,Gefen_PRL_1981} etc.

\noindent Properties of many electron systems on fractal lattices have been studied.  In a detailed work, analytical results were provided for the exact solution of a tight binding model on a fractal \cite{Domany_PRB_1983}, showing that the spectrum also has a self-similar pattern. A recent study has shown that the transmission properties from such a lattice can capture its Hausdorff dimension \cite{Veen_PRB_2016}. Interestingly, the effect of magnetic field was also investigated \cite{Alexander_PRB_1984}. A fractal lattice that has enjoyed continued attention is the ``Sierpinski gasket"  \cite{Fukushima_PA_1992, Wang_PRB_1995, Arunava_JPCM_1996, Pal_PRB_2012, Pal_arXiv_2017} and, has also been realized experimentally \cite{Gordon_PRL_1986, Shang_NatChem_2015}. The present paper is aimed at investigating the question: what kind of topological phases can a fractal host? This question has received scant attention \cite{Song_APL_2014} perhaps due to the fact that the tenets underlying topological phases rely heavily on the notion of a well defined spatial dimension allowing for an unequivocal distinction of bulk and edge.  

\noindent In this paper we investigate several models constructed from fractal structures which allow for variable sharpness in the distinction between the bulk and the edge. Our main finding is that homogeneous fractals, where every site is equally coordinated, do not host gapped topological phases but instead opt for a new metallic phase not usually found in fractal lattices. This metal has intriguing properties for e.g.,~in the Sierpinski gasket fractal, it has excitations which are chiral in nature. This work opens up new direction in the physics of topological phases, pointing to a possible more general classification of topological phases than hitherto available \cite{Kitaev_AIP_2009, Schnyder_PRB_2008}.

\noindent To investigate topological phases in fractals we shall use the Sierpinski gasket\cite{Domany_PRB_1983} as the work horse. The standard Sierpinski gasket is constructed recursively, generation by generation, by starting from  three sites connected by bonds giving a triangular shape. In the next ``generation" sites are added to the midpoints of the bonds (and the length of the bonds is also doubled). Additional new bonds are added such that each new site is four coordinated (i.e.,~has four bonds emanating from it). A thermodynamically large sample can be obtained by repeating this procedure over large number of generations. As is well known, the object thus constructed has a Hausdorff dimension of $\frac{\log 3}{\log 2}=1.58$\cite{Mandelbrot_Book_1983}. An important point to be observed about the above construction is that the first generation sites are only two coordinated while all others are four coordinated. In this sense this system is inhomogeneous. We avoid this by placing the gasket on a torus or sphere. \Fig{SierTB} (top) illustrates the embedding of the gasket on a torus by joining two ``oppositely oriented" gaskets and identifying the first generation sites denoted by $A$ and $B$. Simple tight binding model on this system has already been investigated \cite{Domany_PRB_1983}. In this model the coordination number is 4 for every site. At generation $g$ the number of sites is $3^{g+1}$, and the thermodynamic limit is identified by $g \rightarrow \infty$.  The main finding is that the spectrum has a self similar structure and has infinitely many band gaps in the thermodynamic limit. This should be contrasted with a single band tight binding model on any crystalline Bravais lattice which produces a gapless spectrum. In a simple crystalline lattice of size $L^d$ ($d$ is dimension), the number of bulk sites scale as $N_B \sim L^d$, while the edge sites (sites with lesser neighbors) go as $N_S \sim L^{d-1}$. In thermodynamic limit $\frac{N_S}{N_B}\rightarrow 0$. On the Sierpinski gasket the notion of bulk and edge is not immediately obvious. One natural definition applicable to a generation $g$ is to treat sites of the youngest generation to be the bulk sites while those of all the previous generations to be as edge sites. Thus, for the Sierpinski gasket, $N_B = 2 \times 3^g$ and $N_s = 3^g$ leading to $\frac{N_S}{N_B} = \frac{1}{2}$ (for all $g$).

\begin{figure}
\includegraphics[width=1.0\columnwidth]{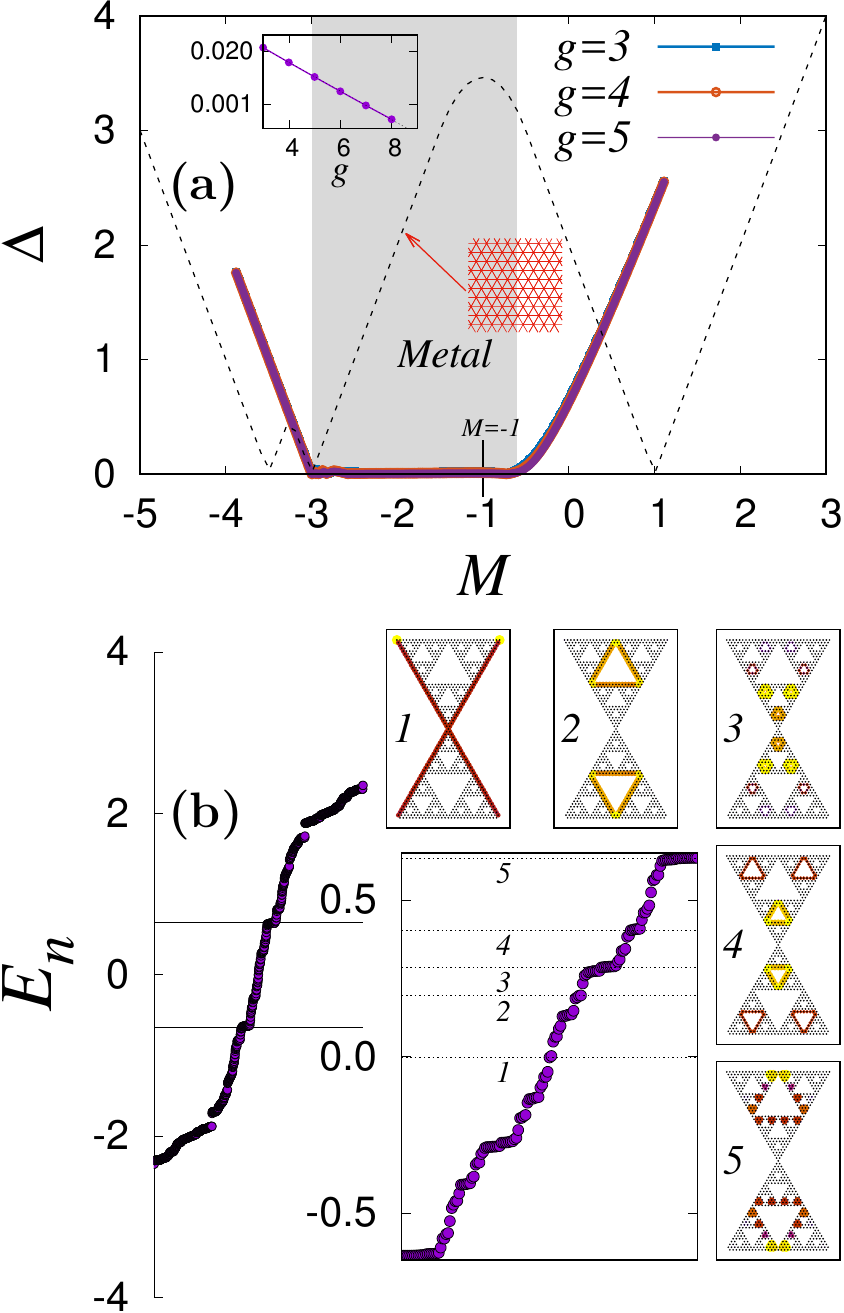}
\caption{{\bf ``Topological" Sierpinski gasket:} (a) Energy gap($\Delta$) at half filling for various generations, shows a vanishing value in range of parameter $M$. Inset shows the gap at $M=-1$ and that it reaches $zero$ with increasing $g$. The dashed line is for the triangular lattice system which has a large finite gap. (b) Left: Energy spectrum at $M=-1$. Right: Zoomed spectrum; insets 1-5 show probability densities of the eigenvectors at energies marked 1-5.}
\label{top}
\end{figure}

\noindent We now construct a topological Hamiltonian on the Sierpinski gasket. For this we consider a two-orbital model on each site, and the Hamiltonian is given by 
\beq\mylabel{eqn:HamGen}
{\cal H} = \sum_{I\alpha} \sum_{J \beta} t_{\alpha\beta}(\br_{IJ})c^\dagger_{I,\alpha}c_{J, \beta},
\eeq 
where $I,J$ are summed on the bonds as shown in \Fig{SierTB}. $c^\dagger_{I \alpha}$ represents a fermion creation operator at site $I$ with a orbital flavor $\alpha$. $\br_{IJ}$ is the vector connecting sites $I$ and $J$.  The hopping matrix  $ t_{\alpha \beta} (\br=\bzero )= $Diag$\{2+M, -(2+M)\}$
and $t_{\alpha \beta}(\br \neq \bzero) =
\begin{pmatrix}
\frac{-1}{2}  & \frac{-ie^{-i\theta}}{2} \\
\frac{-ie^{i\theta}}{2} &   \frac{1}{2}
\end{pmatrix}
$, where $\theta$ stands for the angle made by the bond with the $x$-axis (see \Fig{SierTB}(top)). This is inspired by a similar hopping problem defined on a square lattice \cite{Bernevig_Book_2013}. More pertinent for later comparison is this model defined on a triangular lattice (six coordinated) where one obtains a gapped topological phase in the range $ -\frac{7}{2} < M < 1$ (see SM) when the number of fermions per site is one (half filling). Further, this model has a topological phase even in an amorphous setting \cite{Adhip_PRL_2017}. 

% For both these lattice models, the topological transitions coincide with band gap closings and band inversion. This model was also shown to be topological on an amorphous system \cite{Agarwala_arXiv_2017}. It is important to note that for lattice based systems, the bulk band is continuous over the Brillouin zone; the concepts of Pancharatam-Berry phase and a corresponding nontrivial Chern number, defined as an adiabatic integral over the band Berry curvature is well defined.

\noindent \newline
Retaining a fermion filling of half, \Fig{top} shows the spectrum and states of the ``topological Sierpinski gasket". For values of $M$ with large magnitude, as expected, we find a fully gapped phase. For an intermediate value of $M$ $(-3 \lesssim  M \lesssim -0.5)$, we find that the system becomes gapless (see inset of \Fig{top}(a)). This is to be contrasted with the triangular lattice where one gets a gapped topological phase (see dashed line in  \Fig{top}(a)). The nature of the states in this Sierpinski gasket is shown in \Fig{top}(b), for $M=-1$ (where the triangular lattice has large gap). States near the chemical potential have a remarkable character. They all appear to behave like ``edge states" living on triangular motifs bounded by sites of various generations. The states closest to the zero energy are on triangles bounded by the sites of the earliest generations. The metallicity arises from the fact that there are a large number of states (proportional to the total number of sites) near the chemical potential leading to a finite density of states. Quite interestingly the gapped features found in the single band tight binding model are completely washed out by the ``topological Hamiltonian" and the fractal is rendered metallic! Given the nature of the spatial structure of states near the chemical potential we call this this ``fractalized metal". Interestingly fractalized metal is ``topologically trivial" as evidenced from the fact that the Bott index \cite{Loring_EPL_2010} vanishes.

\begin{figure}
	\includegraphics[width=1.0\columnwidth]{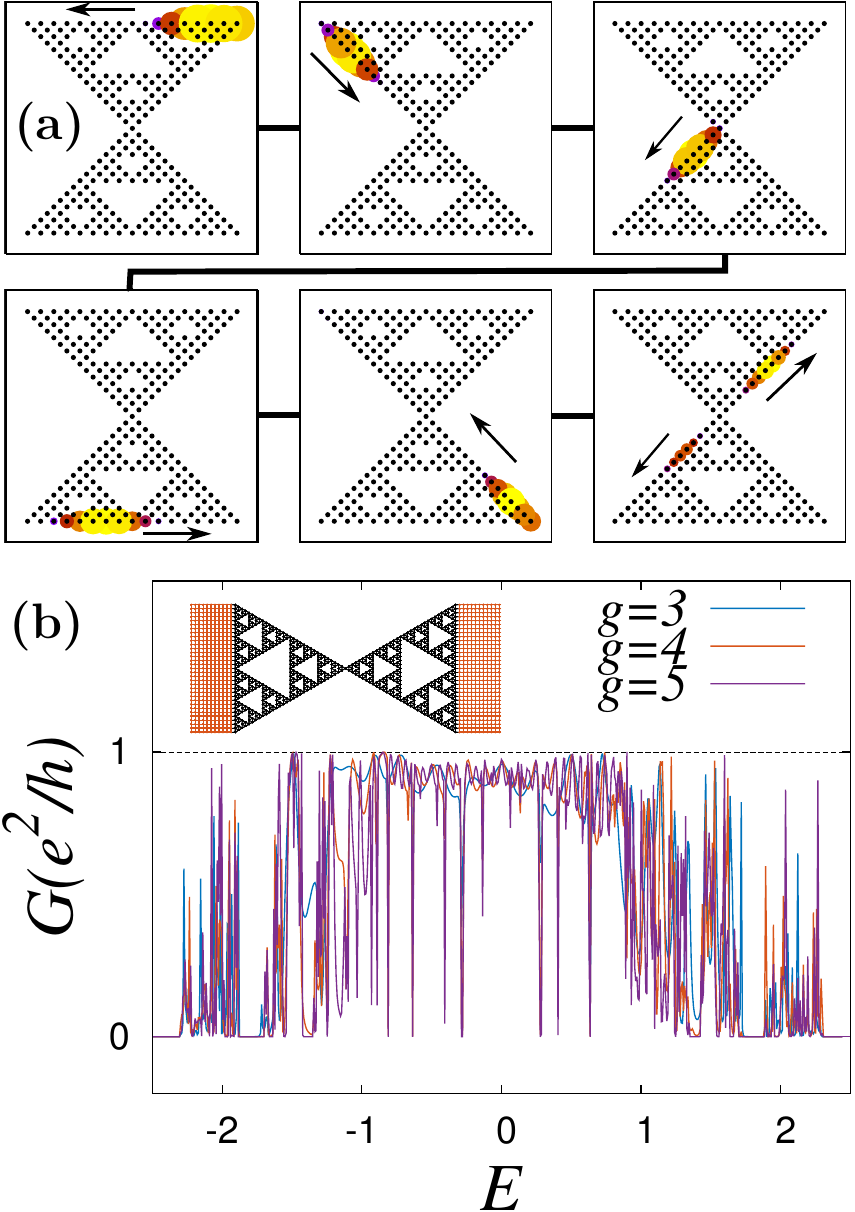}
	\caption{ {\bf Transport:} (a) A wave packet created at the top-right edge of the gasket is projected on the occupied states within the energy range ($-0.1BW<E-E_F<0.1 BW (E_F=0)$). The sequential panels show the evolution of such a packet. It can be seen that it moves exclusively on the edge of the system with a particular chirality. Here $g=4, M=-1$. (b) Two terminal conductance ($G$) as a function of energy ($E$) through the Sierpinski gasket when it is in the gapless regime ($M=-1$).}
	\label{transport}
\end{figure}

\noindent It is interesting to explore the nature of transport in the fractalized metal. To investigate this we employ two methods. We construct an initial state at a corner of the Sierpinski gasket and project it onto the states between the energy($E$) interval $\{-0.1BW<E-E_F< 0.1BW\}$ where $BW$ is the bandwidth. These states are primarily comprised of the ``edge states" of the kind discussed in \Fig{top}(b). The time evolution of such an initial state is shown in \Fig{transport}(a), where we see that the excitations have a distinct chiral character unlike a usual metal. Interestingly when the initial site is chosen to be in one of the ``inner edges", the wavepacket moves in a sense opposite opposite to that on the outermost edge (see SM for details). In fact, this fractalized metallic state is truly unique in that such excitations are not found even in a metal created by partially filling a band that is topologically nontrivial (an example is the half filled valence band of the triangular lattice model discussed above). Further, we have also studied the effect of disorder and found that the chiral motion is not significantly perturbed by weak disorder (Anderson disorder drawn with width $\approx 0.1BW$), pointing to the robustness of this metal. We additionally investigate the transport properties via the Non-Equilibrium Green's Function (NEGF)\cite{Datta_Book_1997} method by connecting the Sierpinski gasket to leads (see \Fig{transport} (b)). Interestingly we do not find conductance quantized to unity; although it is quite close to it for most values of energy (also see SM  for results in presence of disorder). The occasional sharp dips arise from the fractalized nature of the states i.e., low energy states that lie away from the peripheral sites that are in contact with the leads (see for e.g.,~\Fig{top}(b)-4). This is, again, unusual for a typical metal. 

\begin{figure}
	\includegraphics[width=1.0\columnwidth]{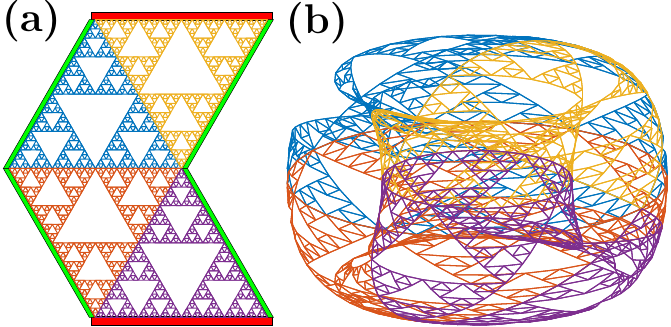}
	\caption{ {\bf Combining Sierpinskis:} (a) Four Sierpinski triangles can be combined in a way that can made into a torus (shown in (b)). The red patches (top and bottom) are glued together, as are green (right and left). This system shows a nontrivial Bott index and produces a topologically gapped phase in an intermediate range of $M$.}
	\label{SierToru}
\end{figure}

\noindent It is natural to ask about the conditions necessary to obtain this fractalized metal. Is a finite $N_S/N_B$ ratio sufficient? We show below that a finite $N_S/N_B$ ratio will not always give a fractalized metal. \Fig{SierToru} shows a fractal model constructed by combining four Sierpinski gaskets along the edges of the largest triangles. Although $N_S/N_B$ is finite in this system, one finds that the system develops a gap (owing to the hybridization of low energy states of the individual triangles) and in fact is topological (in a certain window of $M$) with a nontrivial Bott index! We have also constructed other lattices such as the Sierpinski carpet which also shows gapped topological phases. Based on the above, a necessary criterion we need to obtain the fractalized metal seems to the homogeneity of coordination in the lattice. In Sierpinski gasket all sites are four coordinated while the on the Sierpinski carpet there are a finite fraction of sites with a different coordination. In both cases these different coordinated states can be held responsible for the hybridization and gapping out of the low energy states.

\begin{figure}
	\includegraphics[width=1.0\columnwidth]{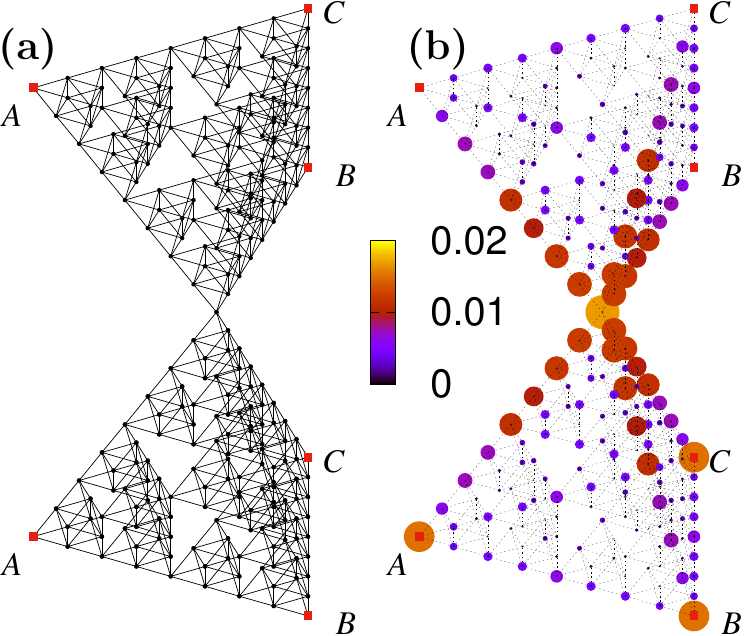}
	\caption{ {\bf Sierpinski Tetrahedrons:} (a) Sierpinski tetrahedron where a tetrahedron replaces the triangular motif of the Sierpinksi gasket. While the system is embedded in three dimensions, its Hausdorff dimension is $2$. The figure shown is for $g=3$. (b) Setting up of a topological Hamiltonian results in the formation of surface states which look like connections made of various ``corner" states.}
	\label{SierTetra}
\end{figure}

\noindent We now illustrate another example of a fractal system which shows intriguing features. We construct a ``three dimensional" version of the Sierpinski gasket where a tetrahedron replaces the triangular motif of the Sierpinski gasket. This object can be embedded on a three torus by identifying sites marked $A,B,C$ in \Fig{SierTetra}. While one might naively expect that this system should be a straightforward generalization of the previous example, this is rather subtly different. The edges in the Sierpinski gasket can be considered  ``one-dimensional" as every inner triangle has a one-dimensional perimeter. The Sierpinksi tetrahedron has infinite number of surfaces, however, each of these surfaces are itself fractals -- Sierpinski gaskets. Setting up a four-orbital model of a topological insulator in this system \cite{Adhip_PRL_2017}, one finds that close to $E=0$ one can have ``surface" state which looks like a network of corner states (see \Fig{SierTetra}). With increasing generation, one also finds that this system is a metal. This same model, in the same parameter regime, is known to produce surface states in a cubic lattice and also on an amorphous system \cite{Adhip_PRL_2017, Fradkin_Book_2013}. It therefore shows that fractalized metallic state is possible even in this system.

\noindent To conclude, we have explored the possibility of topological phases on fractal lattices. This leads us to an interesting conclusion, that homogeneous fractals will host a fractalized metallic phase and not a gapped topological phase. This is demonstrated in two fractals with different dimensions. This work brings a fresh perspective on the phases of noninteracting fermions revisiting the notion of bulk edge correspondence, particularly in a system where these separation is nebulous. An interesting future direction will be to investigate the effect of interactions on such fractalized metals.

\noindent {\bf Acknowledgements:} AA thanks CSIR, India and Max-Planck Grant through the partner group between ICTS and MPIPKS for funding. VBS acknowledges generous support from SERB, DST, India.

%\bibliography{refFrac,RandomTI}

\putbib[fullref]
%\putbib[refFrac,RandomTI]
%\bibliography{RandomTI}
\end{bibunit}

%\ifdefined\makeSM
%%%%%%%%%%%%%%%%%%%%%%%%%%%%%%%%%%%%%%%%%%%5
%\newwrite\tempfile
%\immediate\openout\tempfile=junkSM.\jobname
%\immediate\write\tempfile{\noexpand{\thepage} }
%\immediate\closeout\tempfile

\onecolumngrid

\begin{bibunit}

\relax
\clearpage
\newpage

\appendix

\setcounter{page}{1}
\setcounter{figure}{0}
\setcounter{section}{0}

\renewcommand{\appendixname}{}
\renewcommand{\thesection}{S\arabic{section}}
\renewcommand{\thetable}{S\Roman{table}}
\renewcommand{\figurename}{Supplementary Figure}
\renewcommand{\thefigure}{S\arabic{figure}}
\renewcommand{\theequation}{\thesection.\arabic{equation}}

%\widetext

\centerline{\bf Supplemental Material}
\centerline{\bf for}
\centerline{\bf \mytitle}
\centerline{Adhip Agarwala, Shriya Pai and Vijay B.~Shenoy}
\author{Adhip Agarwala}
\author{Vijay B. Shenoy}
%\email{}
%\email{shenoy@physics.iisc.ernet.in}
%\affiliation{Centre for Condensed Matter Theory, Department of Physics, Indian Institute of Science, Bangalore 560 012, India} 
\vspace{0.5cm}
Here we outline the methods used in this work and present additional results. 
% % % % % % % % % % % % % % % % % % % %
{\centering
\section{Bernevig-Hughes-Zhang (BHZ) model on a triangular lattice}
\label{BHZtri}}

\begin{figure}
\includegraphics[width=0.6\columnwidth]{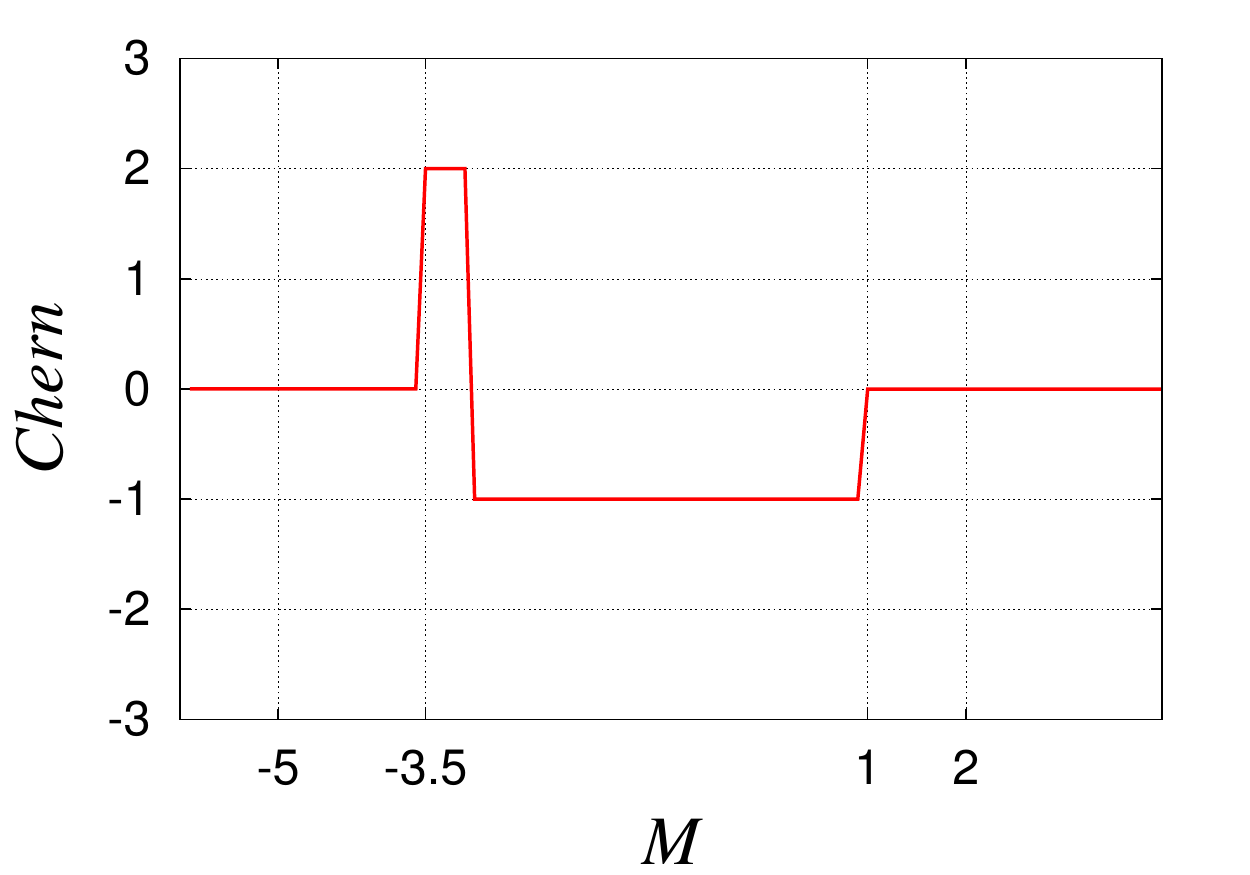}
\caption[BHZ model on a triangular lattice]{{\bf Triangular lattice:} The variation of the Chern number of the lower band (Fermi energy $E_F=0$) as a function of $M$ for the system described by the Hamiltonian (see \eqn{trianHamil}) defined on a triangular lattice.}
\label{trichern}
\end{figure}

We can implement the BHZ model\cite{Bernevig_Book_2013} as described in the main text (see \eqn{eqn:HamGen}) on a triangular lattice. The dispersion is given by

\bea
H &=& \sigma_x\Big(-\sin(k_x)-\sin(\frac{k_x}{2})\cos(\frac{\sqrt{3}}{2}k_y)\Big)+\sigma_y\Big(-\sqrt{3} \cos(\frac{k_x}{2})\sin(\frac{\sqrt{3}}{2}k_y)\Big) \notag \\ &+&  \sigma_z \Big(M+2-[\cos(k_x) + 2 \cos(\frac{k_x}{2})\cos(\frac{\sqrt{3}}{2}k_y)]\Big).
\label{trianHamil}
\eea
where $\sigma$s represent the two-orbital basis on every site. $k_x, k_y$ are the components of momentum vector $\bk$ defined on the Brillouin zone. The variation of the gap as a function of parameter $M$ was shown in \Fig{top}(a). The Chern number variation is shown in \Fig{trichern}. The model (\eqn{eqn:HamGen} in the main text) therefore provides for a topological phase both in square lattice \cite{Bernevig_Book_2013} and on a triangular lattice. 

\vspace{0.5cm}
{\centering
\section{Self-similar edge spectrum in Sierpinski gasket}}

The Sierpinksi gasket which we have discussed in detail in the main text has a self-similar spectrum when every site is coupled via a hopping $-t  (t=1)$ \cite{Domany_PRB_1983}. This self-similarity continues even when a topological model is setup on this fractal. In \Fig{top}, we saw that close to $E=0$ the states tend to reside on select ``edges". Which edge will in general be preferable also follows a self similar pattern which we now discuss.

\begin{figure}
\includegraphics[width=0.55\columnwidth]{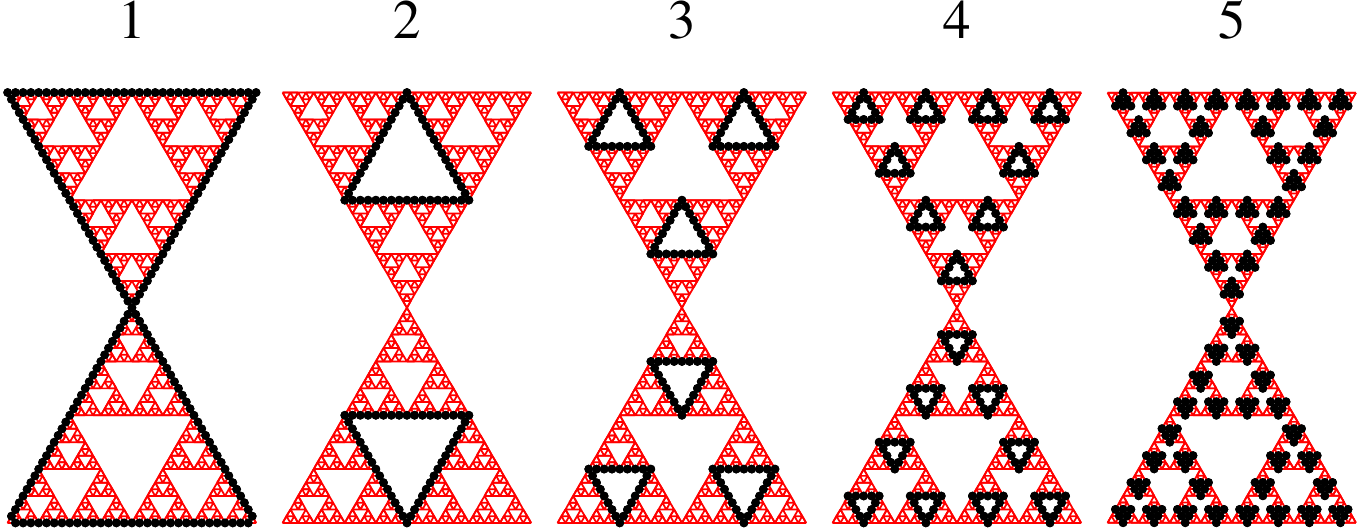}
\includegraphics[width=0.65\columnwidth]{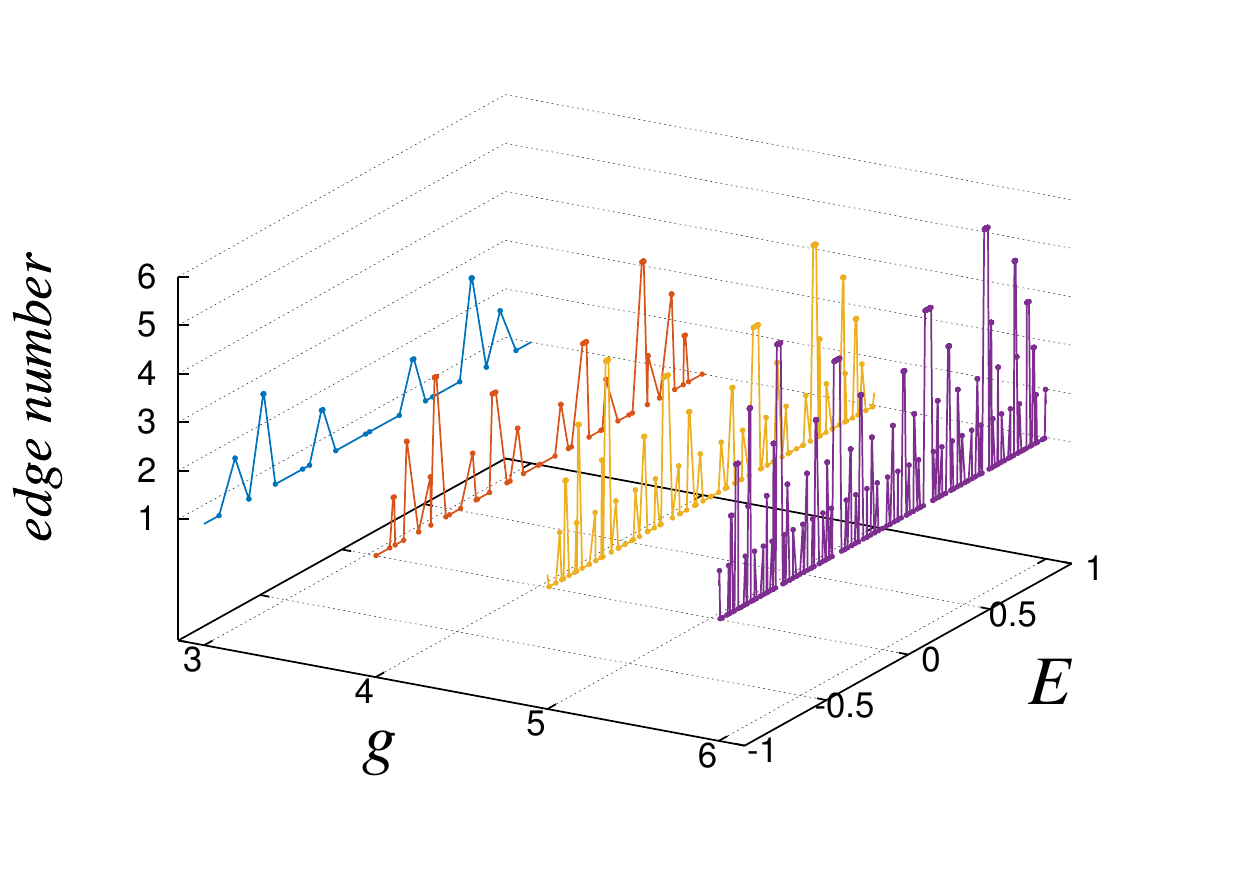}
\caption{{\bf Self-similar edge spectrum:} (Top) The various ``edges" of the Sierpinski gasket is shown ($I=1,\ldots,5$). (Generation($g$)=5) (Bottom) The states and in which edge they lie, as a function of generation number $g$ and energy $E$. These are dominantly ``edge" states and exist at various edges in an interesting self-similar pattern $(M=-1)$.} 
\label{edgesgen}
\end{figure}

In order to analyze the complete edge spectrum we number the edges using $I$, where $I=1$ shows the outermost edge and $I$ progressively increases through integers as one goes deeper into the lattice (an example is shown in \Fig{edgesgen}). For any wavefunction $|\psi \rangle = \sum \psi_i | i\rangle $, one can evaluate the overlap with the edge, 

\beq
O_I =  \sum_{i \in I} |\psi_i|^2 
\eeq
where $i$ is summed over all the sites which belong to the edge $I$. 

Now for any $m$th wavefunction, we define  $O^m= \max\{O_I, {I \in 1, \ldots, m} \}$ which shows the value of the maximum probability of a wavefunction to be an edge state. We can also define $I^m = \{I, I \widehat{=}  O^m\}$ (i.e., the edge number that corresponds to the maximum overlap $O_I$ ). Therefore a state which resides, dominantly on an edge with a value of $O^m$ close to $\sim 1$ demonstrates that the particular wave function resides on that particular edge. The plot of $I^m$ as a function of $E$ and generation number $g$ is plotted in \Fig{edgesgen} for $M=-1$. The values of $O_I \sim 1$ for all these states. The self similarity in the spectrum as a function of generation is manifest.

%One can calculate $O^m$ and $I^m$ as a function of $E$ and $M$, the results are shown in \Fig{edgespec4} and \Fig{edgespec5}. The self-similarity between the spectrum in the two generations are clear. Also the zooming the closer to $E=0$ region, for $5$th generation is strikingly similar to the same for $4$th generation. As can be seen with increasing generation since the number of edges also increase, in the infinite generation limit, we have a infinite edge spectrum which is self similar. This, to our knowledge, is the first explicit demonstration of topological self similar edge spectrum on a manifestly fractal lattice.

\vspace{0.5cm}
{\centering
\section{Density of states and thermodynamic gaps}}

\begin{figure}
	\includegraphics[width=0.65\columnwidth]{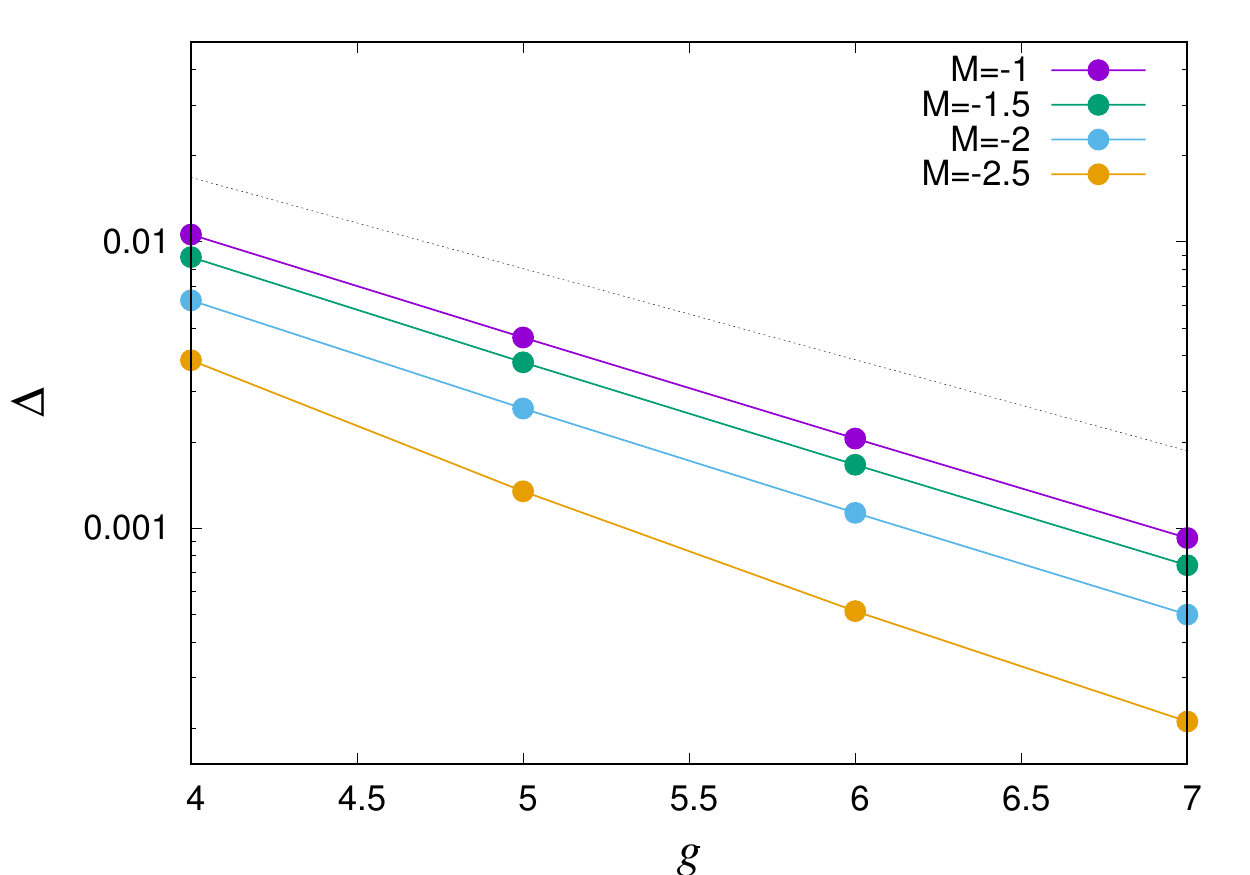}
	\caption[Gap at the fermi energy]{{\bf Scaling of the gap:} The value of gap to excitation at Fermi Energy ($E_F=0$) as a function of generation number $g$ for different for $M$s. The dashed line scales as $\approx e^{-\alpha  g}$ where $\alpha \sim 0.8$. Clearly the system becomes gapless in the thermodynamic limit. }
	\label{GapatMm1}
\end{figure}

\begin{figure}
	\includegraphics[width=0.45\columnwidth]{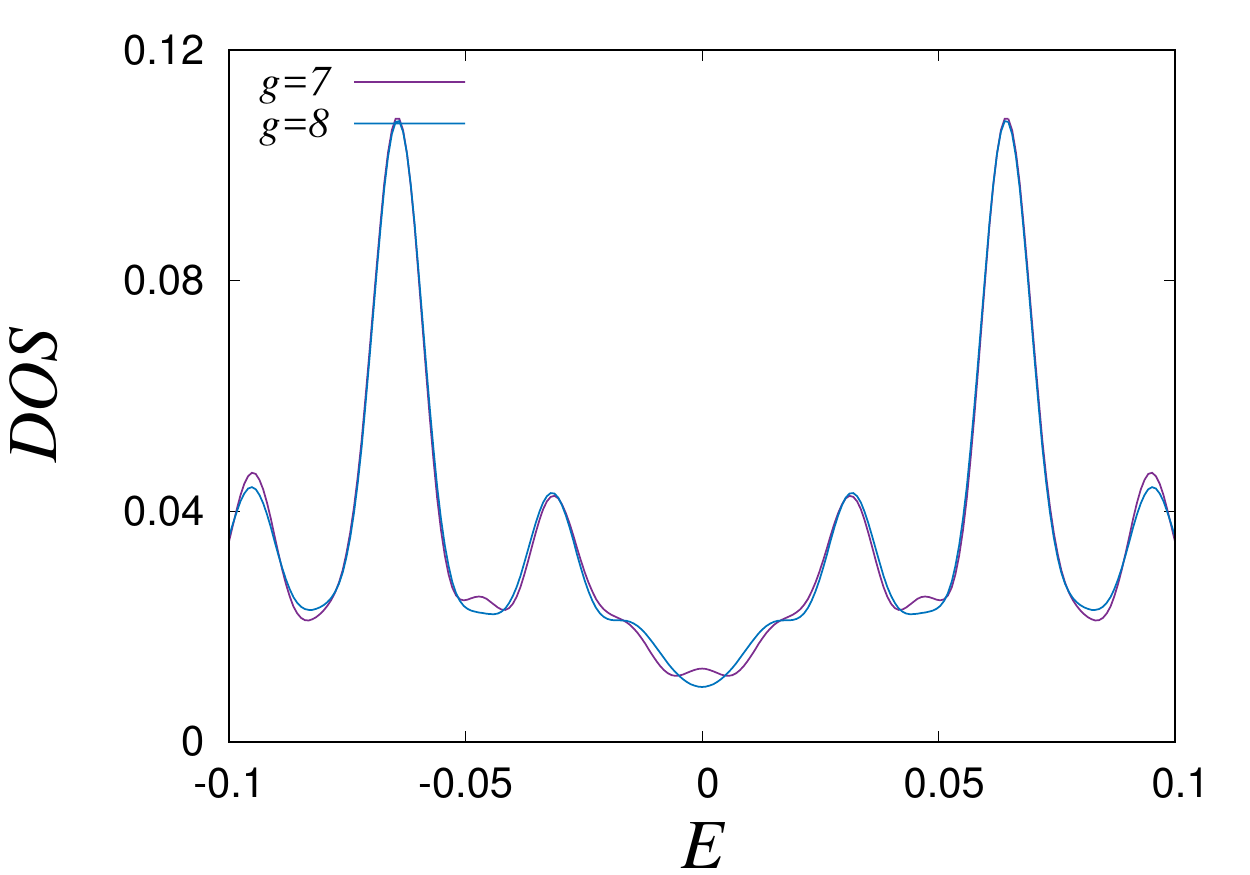}
	\includegraphics[width=0.45\columnwidth]{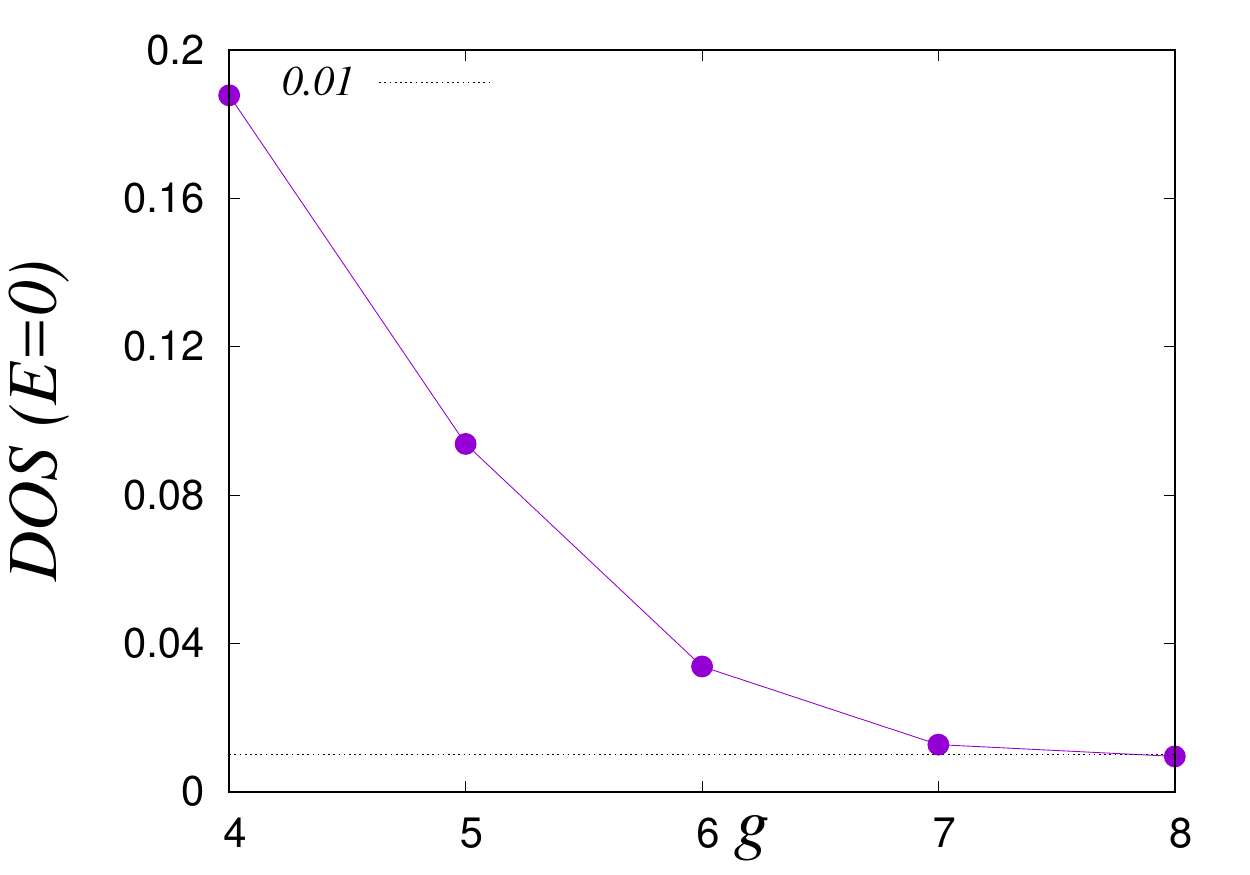}
	\caption[Gap at the fermi energy]{{\bf Density of states:}  (Left) The density of states as a function generation number near $E=0$ for $M=-1$. A broadening Gaussian delta function is used with width$=0.0075$. (Right) The dos value at $E=0$ with generation number. With increasing generation number the dos seems to saturate. Further system size scaling may be necessary to see this more carefully.}
	\label{dosval}
\end{figure}

In \Fig{top} we showed that the Sierpinksi gasket shows a metallic phase in half-filling as a function of $M$.  In \Fig{GapatMm1} we show the scaling of the gap (at half filling) with increasing $g$ for various values of $M$. One finds that the gap goes to zero exponentially with increase in $g$. Also the density of states (close to $E_F=0$) is shown in \Fig{dosval}. DOS seems to reach a finite value with increasing generation $g$. Away from the Fermi energy,  it is interesting to note what happens to ``bulk" bands in this system. In crystalline lattices, for this model, their is a single bulk band for $E>0$ (and symmetrically placed $E<0$ band). In the case of Sierpinksi gasket the system has thermodynamic gaps in the bands which are non-topological in nature. This can be considered as remnants of the finite gaps which occur even when simple tight binding model is implemented on Sierpinski gasket (see \Fig{SierTB}).

\begin{figure}
\includegraphics[width=0.65\columnwidth]{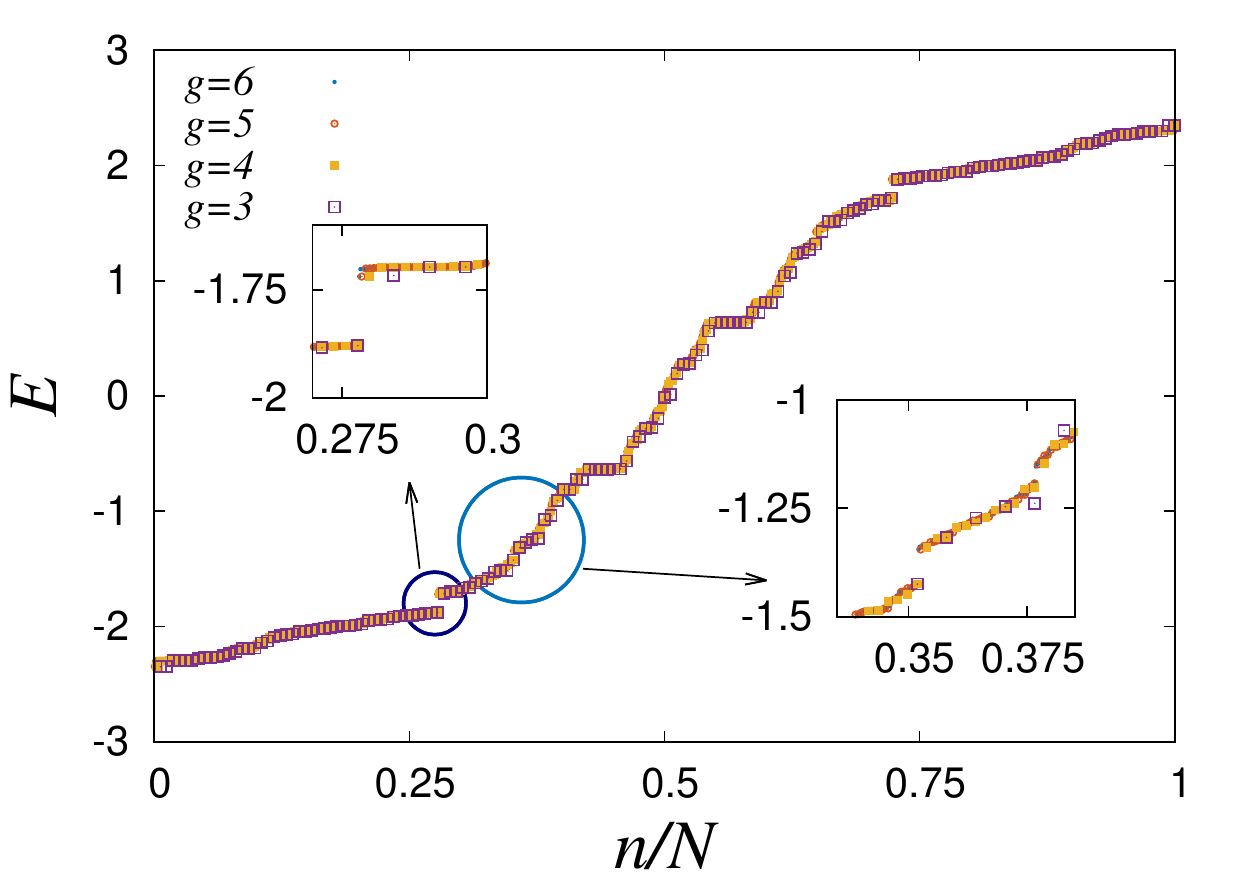}
\caption[Gaps in the bulk states]{{\bf Thermodynamic gaps in bulk states:} The spectrum for the Sierpinksi gasket when the topological Hamiltonian is implemented on it. Results for different generations are plotted for $M=-1$. These gaps are however non-topological (Bott index is $zero$).}
\label{fracgaps}
\end{figure}

%{\centering
%\section{Bott index}}

\vspace{0.5cm}
{\centering
\section{Sierpinski carpet and Torus; Bott index}}

\begin{figure}
\includegraphics[width=0.8\columnwidth]{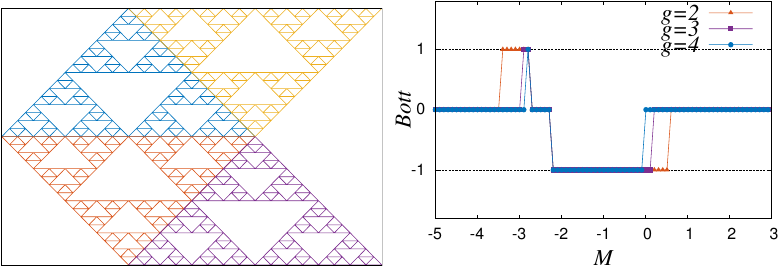}
\caption[]{{\bf Sierpinksi Torus: }(Left) Four Sierpinski triangles can be combined to make a structure which can be made into a torus. (Right) The variation of Bott index, when calculated at half filling, is shown as a function of $M$. One finds a topologically nontrivial regime.}
\label{SiertoBott}
\end{figure}

We now consider two more fractal systems, but where every site is not equivalently coordinated. We set up the same topological Hamiltonian as shown in the main text and calculate the topological index as a function of the parameter $M$. For the first system, we combine four Sierpinski triangles into the form as shown in \Fig{SiertoBott}. One can notice here that the sites belonging to the boundary of the triangles have a larger coordination number. The variation of the Bott index is also shown at half filling. One finds that in this system, under periodic boundary conditions, edge states does not appear at the outermost edge and the Bott index is nontrivial in a regime of $M$.

\begin{figure}
\includegraphics[width=0.8\columnwidth]{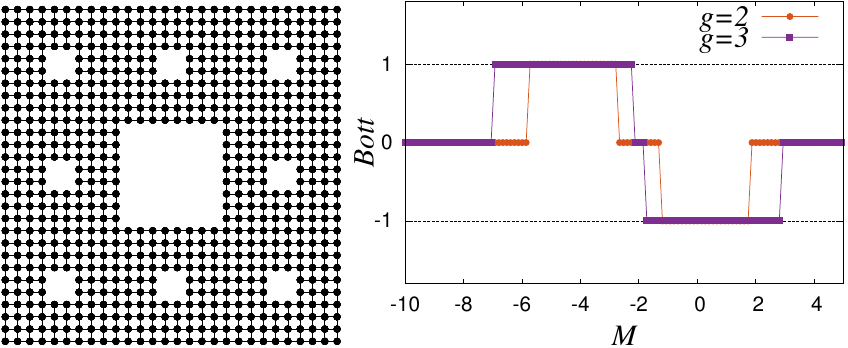}
\caption{{\bf Sierpinksi carpet: } (Left) Sierpinksi carpet is a recursively scooped out square lattice. The picture shown is for a third generation system. (Right) Under periodic boundary conditions, Bott index can be calculated as a function of $M$. The system shows a nontrivial topological regime. }
\label{SierCarp}
\end{figure}

The second system which we analyze is the Sierpinski carpet. The system is shown in \Fig{SierCarp}. Here $1/8th$ of the carpet is scooped out recursively. The Hausdorff dimension for this system is 1.8928. We again set up the same topological Hamiltonian on this system, and find the Bott index as a function of the parameter $M$.
Here again every site is not equivalently coordinated. Another crucial difference between Sierpinski carpet and gasket is the concept of ramification. Ramification \cite{Mandelbrot_Book_1983} counts the number of distinct bonds which need to be deleted to break the fractal into macroscopic objects. For Sierpinski gasket this number is $4$, while for the Sierpinski carpet this number is infinity. It will be interesting to explore, how ramification affects topological phases.

\vspace{0.5cm}
{\centering
\section{Chiral states; Inner edges}}

\begin{figure}
\includegraphics[width=0.8\columnwidth]{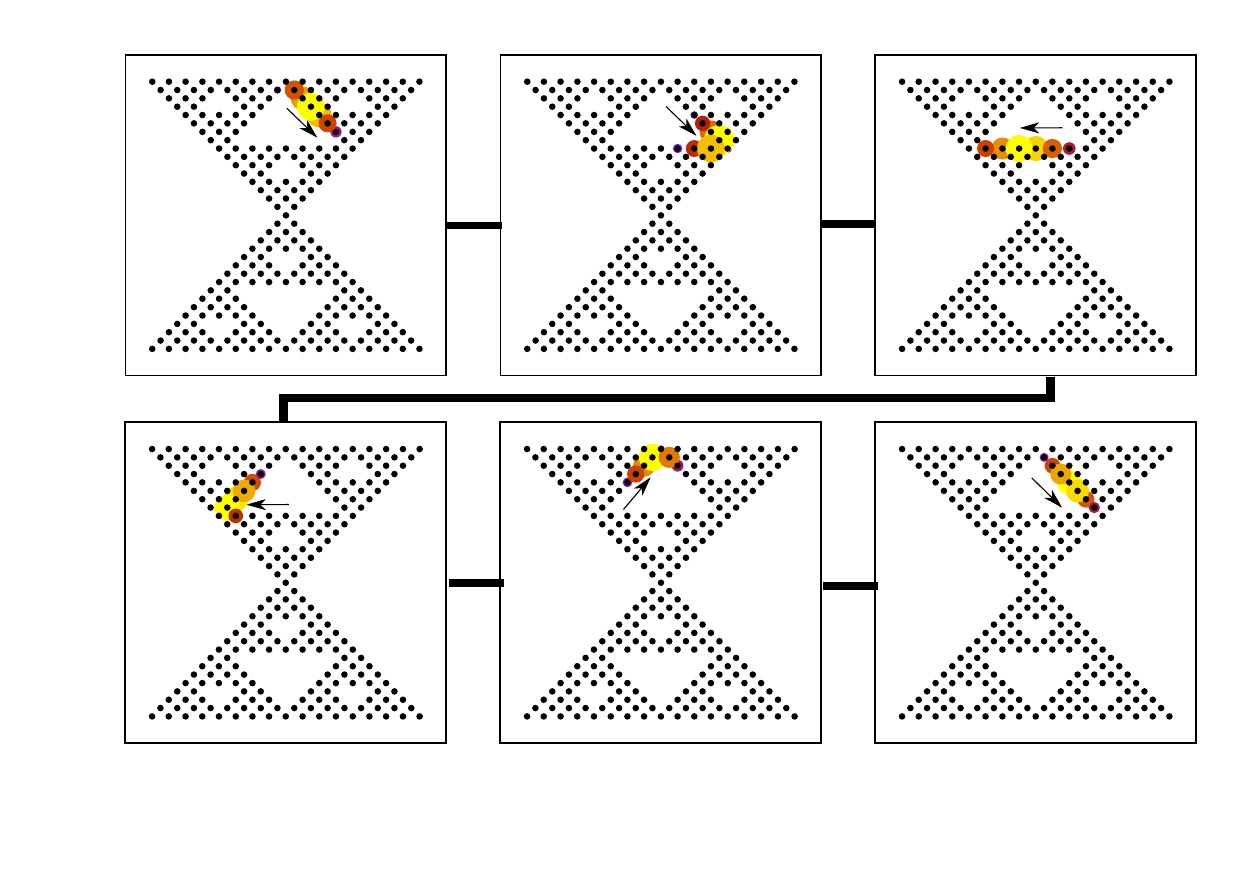}
\caption{ {\bf Wave packet motion in the ``inner edge":} (a) A wave packet created at the top inner edge of the gasket is projected on the occupied states within the energy range ($-0.1BW<E-E_F<0.1BW (E_F=0)$) where $BW$ is the bandwidth. The sequential panels show the evolution of such a packet. It can be seen that it moves exclusively on the edge of the system with a particular chirality. Here $g=4, M=-1$. Interestingly the chirality is opposite to the one seen in \Fig{transport}.}
\label{inneredge}
\end{figure}

In \Fig{transport} we looked at the way in which a state when localized at the outer edge of the system evolves in time. We now analyze the same for an inner edge. The result for the same set of parameters as in \Fig{transport} is shown in \Fig{inneredge}. As can be seen, for an inner edge the sense of chirality is opposite to the case when the wavefunction is localized on the outermost edge.

\vspace{0.5cm}
{\centering
\section{Two terminal conductance and robustness to disorder}}

\begin{figure}
	\includegraphics[width=0.45\columnwidth]{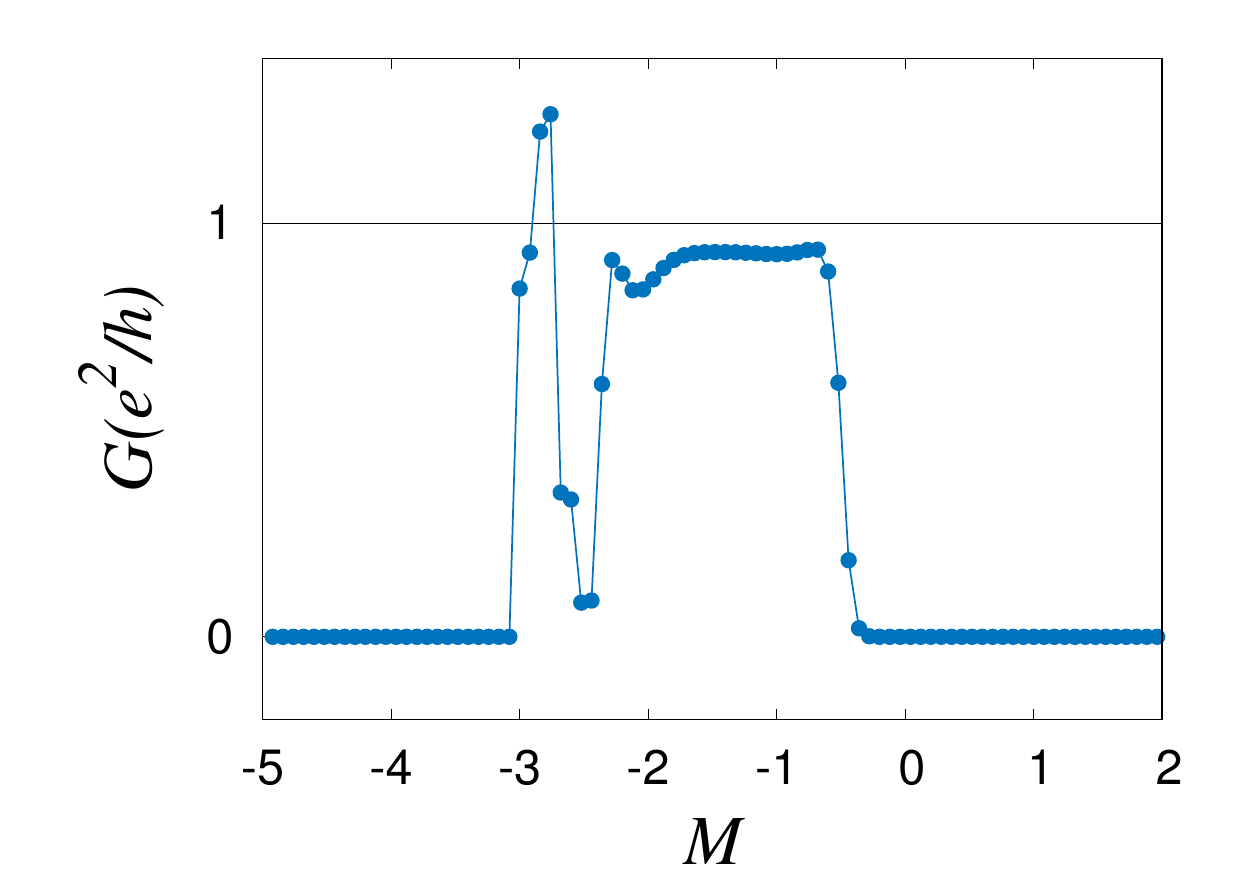}
		\includegraphics[width=0.45\columnwidth]{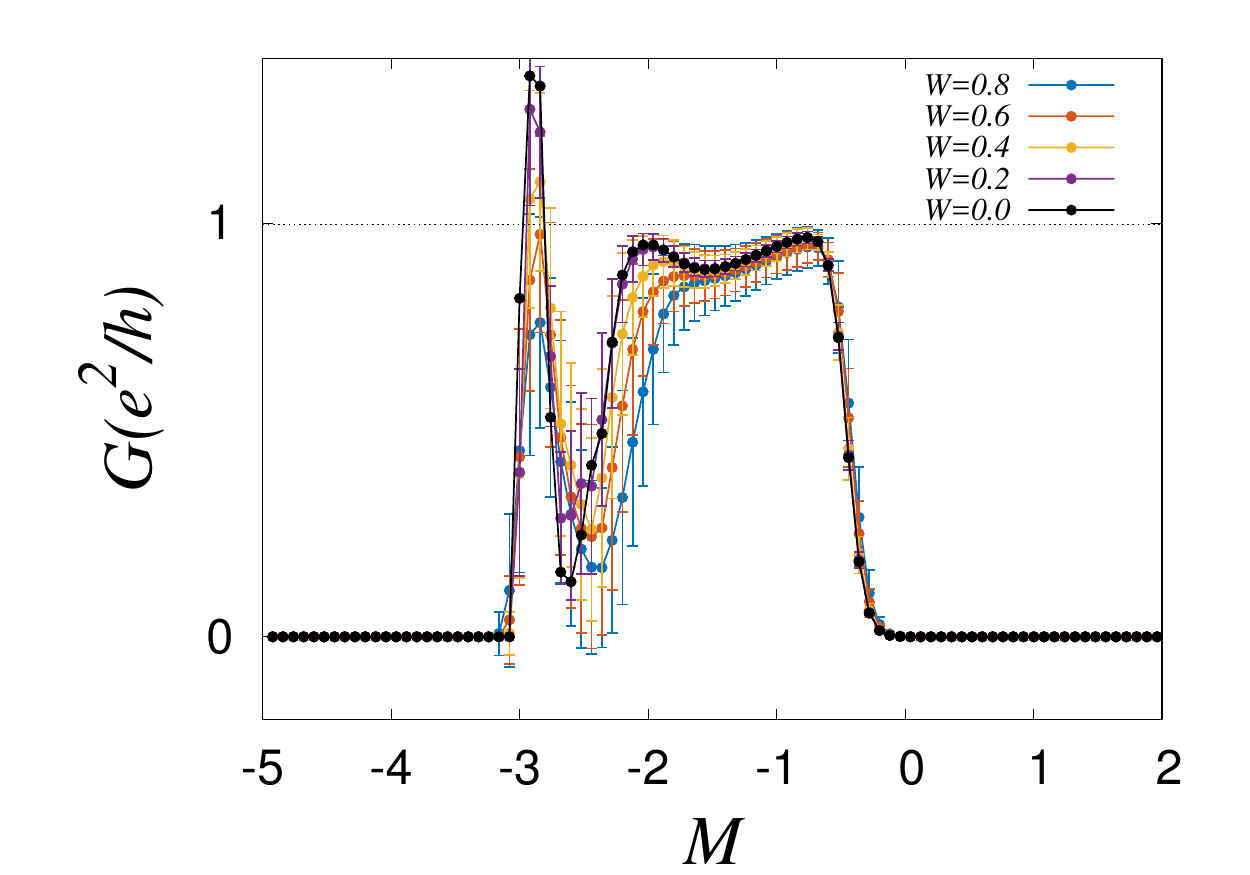}
	\caption{ {\bf Transport and disorder:} (Left) Variation of two terminal conductance $G$ as a function of $M$ when $E_F=0$ and in absence of disorder ($g=4$). The setup is same as shown in \Fig{transport}(b) of the main text.  Conductance is close to $e^2/h$, albeit not exactly. (Right) Variation of disorder averaged conductance (at half filling) with $M$ shows that metallic behavior is robust against weak disorder. Results are shown for $g=3$.}
	\label{GvsMnodis}
\end{figure}

In \Fig{transport}(b) we had looked at the two terminal conductance of the fractalized metal at $M=-1$ and found the conductance to be close to $e^2/h$ near the Fermi energy $E_F=0$. We now scan the value of conductance (fixing the $E_F=0$) and vary $M$.  This variation is shown in \Fig{GvsMnodis}(Left). We now analyze the effect of disorder by adding an onsite potential $\epsilon_{i\alpha}$ to every $i$th site with the orbital label $\alpha$.  $\epsilon_{i\alpha}$ are drawn from a uniform box distribution between $[-\frac{W}{2},\frac{W}{2}]$. Even in presence of a finite $W$, the chiral nature of a wavepacket's time evolution is not significantly perturbed. We now analyze the effect of disorder on the two terminal conductance. The results are shown in \Fig{GvsMnodis}(Right). We find that fractalized metal phase is stable to disorder.   

% % % % % % % % % % % % % % % % % % % % % %
% % % % % % % % % % % % % % % % % % % % % % %

%\putbib[refFrac,RandomTI]
\end{bibunit}
%\bibliography{RandomTI}

%\fi

\end{document}